 \documentclass[final,5p,times,twocolumn]{elsarticle}

 \usepackage{graphics}
 \usepackage{graphicx}
 \usepackage{epsfig}

\usepackage{amssymb}
\usepackage{textcomp}
\usepackage{booktabs}
\usepackage{amssymb,bm,mathrsfs,bbm,amscd}
\usepackage[tbtags]{amsmath}

\usepackage[footnotesize]{caption2}

\usepackage{lineno}

\journal {Nuclear Instruments and Methods in Physics Research Section A}

\begin{document}

\begin{frontmatter}



\title{Study on single-channel signals of water Cherenkov detector array for the LHAASO}

\author[nku]{H.C. Li\corref{author}}
\ead{lihuicai@ihep.ac.cn}
\cortext[author]{Corresponding author. Tel.: +86 010 88236138}
\author[ihep]{Z.G. Yao}
\author[ihep]{M.J. Chen}
\author[nku]{C.X. Yu}
\author[ihep]{M. Zha}
\author[ihep]{H.R. Wu}
\author[ihep]{B. Gao}
\author[ihep]{X.J. Wang}
\author[nku]{J.Y. Liu}
\author[nku]{W.Y. Liao}
\author[ihep]{D.Z. Huang}
\author{for the LHAASO collaboration}
\address[nku]{University of Nankai, Tianjin 300071, China}
\address[ihep]{Institute of High Energy Physics, Chinese Academy of Sciences, Beijing, 100049, China}

\begin{abstract}
The Large High Altitude Air Shower Observatory (LHAASO) is planned to be built
at Daocheng, Sichuan Province, China. The water Cherenkov detector array (WCDA), with an
area of 78,000~$\rm m^{2}$ and capacity of 350,000~tons of purified water, is one of the major components of the LHAASO project.
A 9-cell detector  prototype array has been built at the Yangbajing site, Tibet,
China to comprehensively understand the water Cherenkov technique and investigate the engineering issues of WCDA. In this paper, the rate and charge distribution of single-channel signals are evaluated using a full detail Monte Carlo simulation. The results are discussed and compared with the prototype array.
\end{abstract}


\end{frontmatter}


\section{Introduction}
In very-high-energy (VHE) gamma-ray astronomy, the water Cherenkov technique has the unique advantage of background rejection power superior to those of other ground particle detectors such as plastic scintillators and RPCs. This characteristic has  been well demonstrated by simulations and in practice by the Milagro experiment. New generation facilities, such as HAWC~\cite{DeYoung2012} and LHAASO~\cite{cao2014}, that adopt this technique and with larger area can achieve a sensitivity of more than one order of magnitude better than current experiments.

\begin{figure}
  \centering\includegraphics[width=0.7\linewidth]{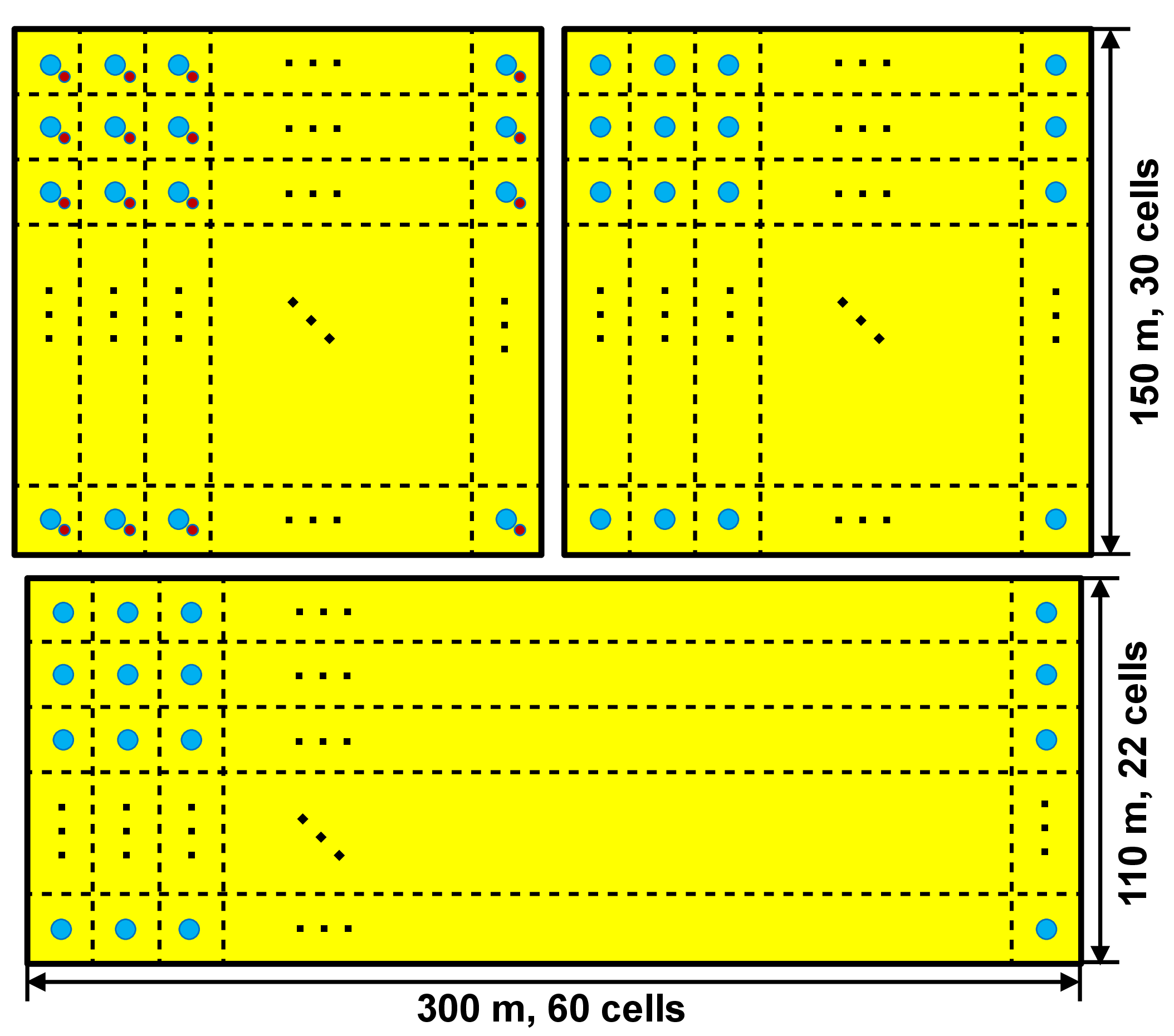}
  \caption{Schematic drawing of the WCDA layout~\cite{li2017}.}
  \label{fig:wcdalayout}
\end{figure}

A water Cherenkov detector array (WCDA) is planned to be built in 2017 at Mountain Haizishan (altitude, 4410 m asl), Daocheng, Sichuan Province, China. This array will cover an area of 78,000~$\rm m^2$ and contain 350,000~tons of purified water. The WCDA will be divided into 3120 detector cells. The main purpose
of the WCDA will primarily survey the northern sky for sources of VHE gamma ray. The whole WCDA will consist of three tanks, two of which will cover areas of
150~m $\times$ 150~m, and the other area of 300~m $\times$ 110~m
(Fig.~\ref{fig:wcdalayout})~\cite{li2017}.  The water in each tank will be subdivided
into cells with areas of 5~m $\times$ 5~m, portioned by black plastic
curtains to prevent the cross-talk of lights between cells.
Additionally, 1 or 2 PMTs will be placed at the bottom of each cell. The PMTs will face upwards with effective water depth 4~m above the photo-cathode.

In the LHAASO-WCDA experiment, to gain a wider dynamic range, the single-channel signals produced by each PMT will be read out via a dynode and anode. A triggerless mechanism for data collection will be adopted. In this mechanism, the single-channel data of every PMT channel will be transferred to a computer cluster for further processing, such as filling histograms, forming triggers, reconstructing online, filtering noise and observing gamma ray burst with low multiplicity studies~\cite{wu2015}.

In single-channel mode, the PMT hits mostly comprise of low photoelectron (PE) signals. These signals are mainly contributed by low-energy (below 100 GeV) cosmic ray showers, major secondary components are photons, electrons (including positrons), muons, neutrons, and protons. The single-channel rate depends on composition, power law spectral indices, interaction model, threshold energy of the primary cosmic rays, water transparency, and effective water depth in the tank.

The single-rate study evaluates the amount of data collection ability and trigger scheme designed. High single-channel rate and cosmic muon influence the event build and event reconstruction. Thus, gaining extensive knowledge of the water Cherenkov detector and determining the accurate noise rates are necessary.

In Section~2, the WCDA prototype array and its data collection modes are introduced. In Section~3, the details of simulations is described. In Section~4, the single-channel rates in the single channel signals and effect factors are analyzed. Finally the study is summarized in Section~5.

\section{WCDA prototype array}\label{sect:ptarray}

 A prototype array  has been built at Yangbajing in 2010. The prototype array~\cite{an2013} has already been operational from 2011 to 2013 for approximately three years. The effective dimensions of the tank are $15~\rm m \times 15~\rm m$ at the bottom, with the tank wall concreted upward along a slope of $45^\circ$ until $5~\rm m$ in height. Thus, the array has an area of $25~\rm m \times 25~\rm m$ at the top. The whole tank is partitioned by black curtains ($4.5~\rm m \times 3~\rm m$) into $3 \times 3$ cells, each one is 5~m $\times$ 5~m in size. A PMT is deployed at the bottom-center of each cell, facing upwards to collect the Cherenkov photons generated by air shower particles in water.

The data collection of the prototype array is specially designed to be multi-purpose to facilitate different detector performance studies. Several data collection modes (trigger modes) are performed repeatedly in turn, with the following settings:
1) single-channel signals with approximately 1/3 PE threshold. These are taken approximately 40 times per day for every individual PMT channel (while other channels are masked), each lasting by approximately 40 seconds;
2) signals of any one of the channels passing with a high threshold (approximately 20 PEs).  These are taken approximately 8 times per day, each lasting for approximately 30 minutes;
3) shower mode, which requires at least 3 PMTs firing at a low threshold during any 100~ns time window;
4) other custom modes for particular analyses such as checking the electronics and testing the time calibration system. The first three components are related to our current study. Combined with the data from the first two modes, the single-channel charge distribution with wide dynamic range for each PMT can be exactly derived. Thus, in this study, the two modes would be unified and assigned as  {\sl the single-channel mode}.

\section{Details of simulations}\label{sect:simulations}

\subsection{Shower simulation}
In this simulation, the air shower events are
generated by CORSIKA v75000~\cite{corsika}. The QGSJET-II
model~\cite{Ostapchenko2006} and FLUKA libraries are used for high energy hadronic interactions and for interaction cross-section in low energy regions, respectively. The loss of the shower information is avoided by setting the kinetic energy cut to lower values than Cherenkov production threshold in the water for secondary particles in CORSIKA, that is, 50 MeV for hadrons and muons and 0.3 MeV for pions, photons and electrons~\cite{li2014}. The detector is supposed to be built at an altitude of 4300 m asl, and geomagnetic field components are set to 34.5~$\rm{\mu}$T and 35.0~$\rm{\mu}$T for north and downward vertical components of geomagnetic field respectively.
Five different primary cosmic ray nuclei (proton/helium/CNO/MgAlSi/Fe) are used in simulation. The primary energy is sampled in the range from GeV to 10~TeV, divided into several different energy ranges, at zenith angles $0^\circ$ -- $70^\circ$, and uniformed azimuth angle $0^\circ$ -- $360^\circ$. For P/He, the fluxes and spectral index measured by AMS-02~\cite{Aguilar1,Aguilar2} are used under 1~TeV/2~TeV, whereas, the spectral index of Horandel model~\cite{Horandel2003} is used. For CNO/MgAlSi/Fe, the fluxes and spectra index  measured by CREAM-II~\cite{Ahn2009} are used. Table~\ref{tab:corsika} lists two important CORSIKA input parameters, different energy range  and  corresponding spectral index, used in the simulation.
\begin{table}[htb]
  \centering
  \caption{Two important parameters of the five primary particles in simulation.}
  \label{tab:corsika}
  \small
  \begin{tabular}{lcccc}
    \hline
    CR & energy range (GeV) & spectral index \\   
    \hline
    P  &1.4 -- 4.1 -- 10.1 -- $10^3$ -- $10^4$  &-1.6, -2.4, -2.7, -2.71  \\
    He      &5.4 -- 10.5 -- 22.3 -- 2$\times$$10^3$ -- $10^4$ &-1.8, -2.3, -2.6, -2.64  \\
    CNO     &17.6 -- 52.6 -- $10^4$& -1.9, -2.60 \\
    MgAlSi  &41.7 -- 200 -- $10^4$ & -2.0, -2.66 \\
    Fe      &97.4 -- 300 -- $10^4$ & -2.2, -2.63 \\
    \hline
  \end{tabular}
\end{table}

\begin{figure}[htb]
  \centering
  \includegraphics[width=0.9\linewidth]{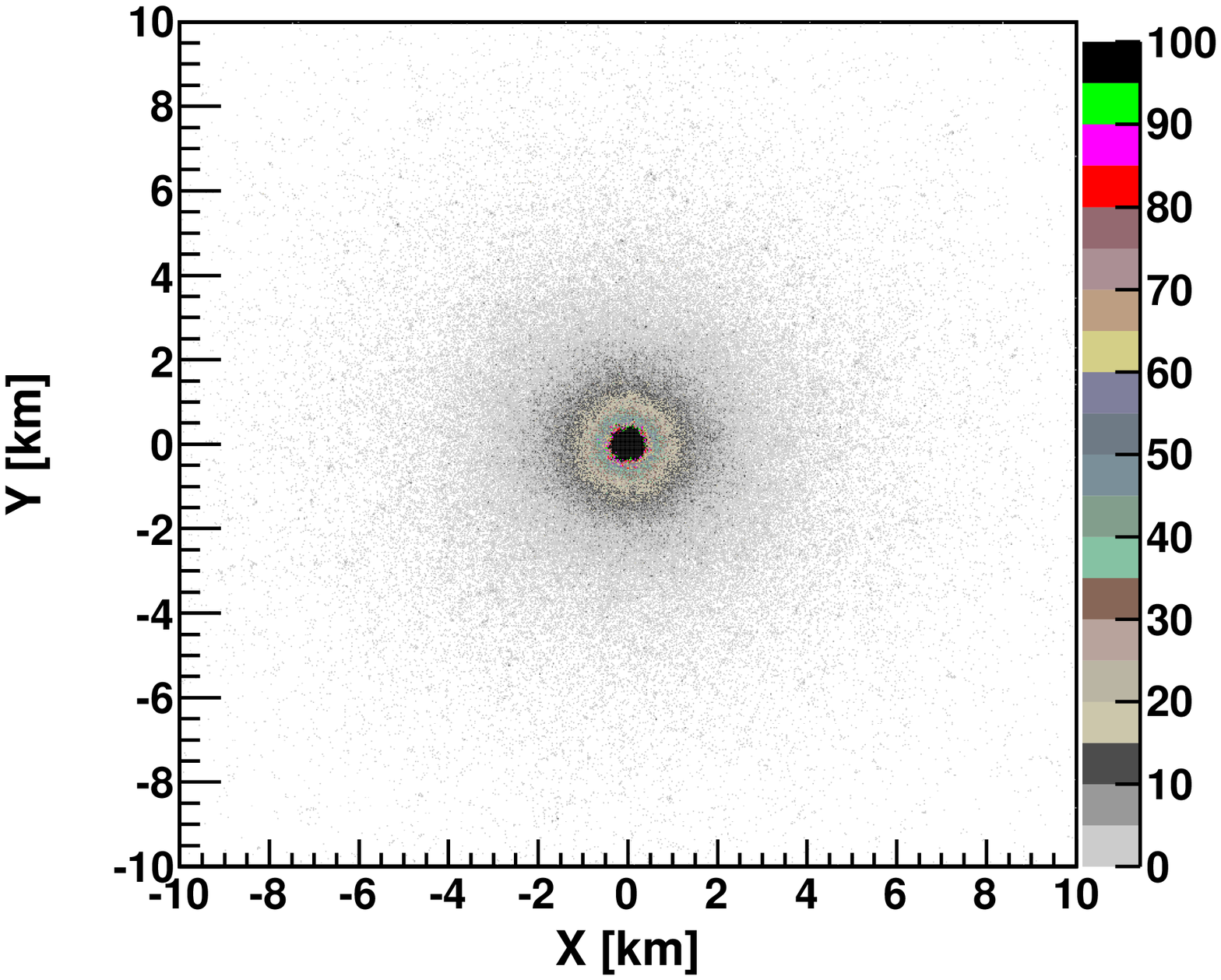}
  \includegraphics[width=0.9\linewidth]{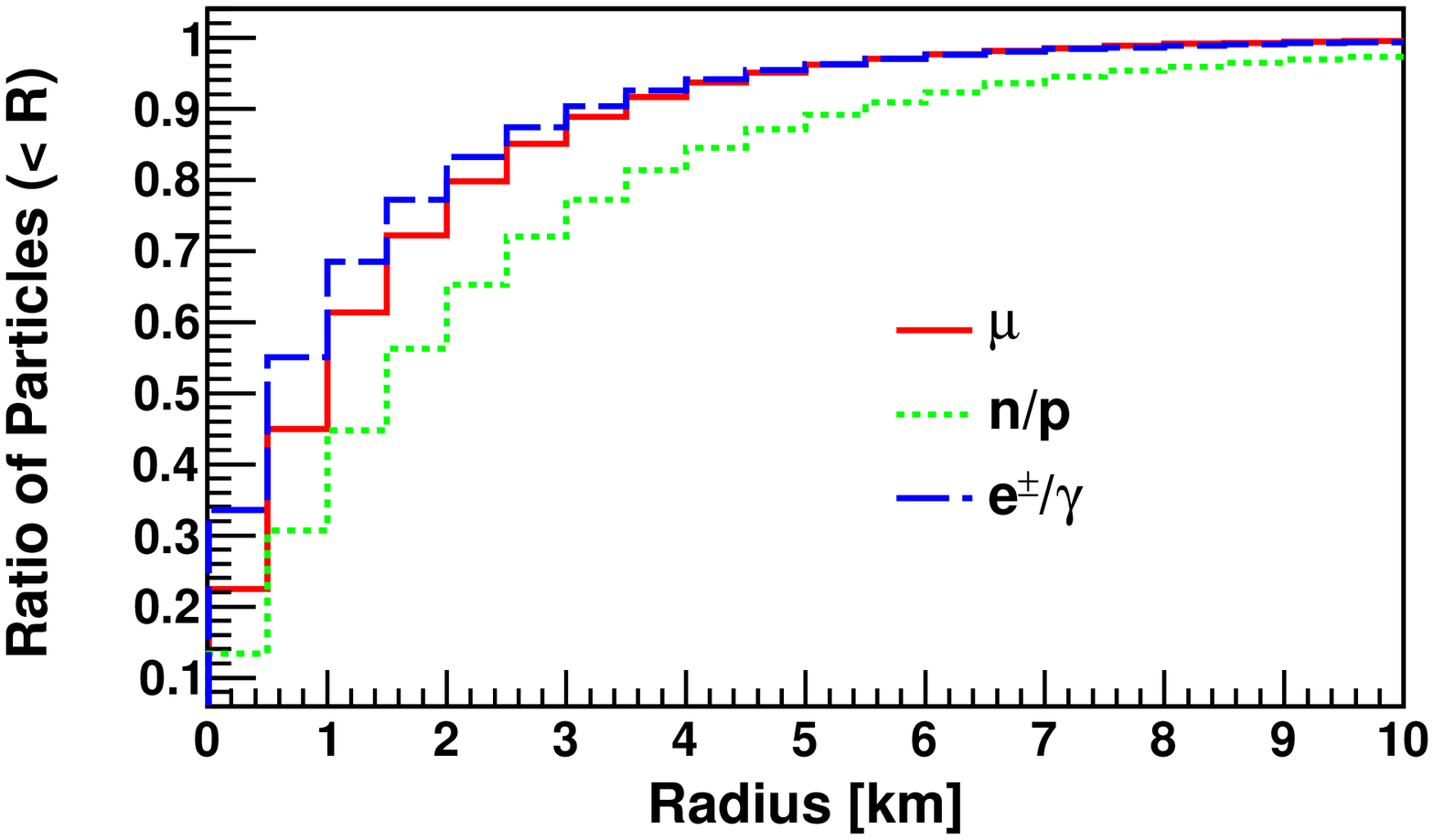}
  \caption{Distribution of secondary particles (top) and the ratio of particles versus the radius (bottom).}
  \label{fig:zcorsikasecall}
\end{figure}

Fig.~\ref{fig:zcorsikasecall} shows the distribution of secondary particles from $10^4$ proton showers, and the energy range is from 10 GeV to 1 TeV. When the collection area reaches $\rm 20\;km\times 20\;km$, the collection rates increases to 97.0\% for neutron and proton (n-p) and 99.5\% for muon and electromagnetic (EM: $e^{\pm}$ and $\gamma$) components.

Detailed simulation is the best approach for calculating the single-channel rate, but in
practice, this process would be exhausting. Simulating shower would involve a very huge amount of data and highly time consuming especially the tracing of Cherenkov lights in the water. A very simple and efficient optimization procedure is adopted to resolve the identified
obstacle and improve the shower utilization ratio. For each shower, secondary particles are collected in the area of $\rm 20\;km\times 20\;km$ to reduce the burden of iterations during simulation. This area is partitioned into $800 \times 800$ grids, each of which is $25~\rm m \times 25~\rm m$ in size corresponding to the prototype array. The secondary particles in one grid are treated  as a new primary shower part, which can reach the detector array. Thus, every shower is reused many times in the simulation.

\subsection{GEANT4 simulations}
The GEANT4 toolkit with 9.1.p01~\cite{Agostinelli2003} version is employed to track the  secondary particles of shower and their productions in the detector, where the PMT models are taken from GenericLAND software library~\cite{neutrino}.

The detector efficiency is a crucial factor for accurate calibration of the detector. When a secondary particle goes through one detector cell, the particle will interact with water and produce much Cherenkov light, some of which may arrive at the PMTs. This then produces electric signals, including time and charge information, those are useful for reconstructing the shower. The absorption or scattering of the photons by the water molecules or impurities in the water plays an important role, because the Cherenkov photons have to travel a certain distance before hitting the PMTs. Thus, the water transparency and quantum efficiency of PMT are two critical factors for the Monte Carlo (MC) simulation.

\begin{figure}[htb]
  \centering
  \includegraphics[width=0.9\linewidth,clip]{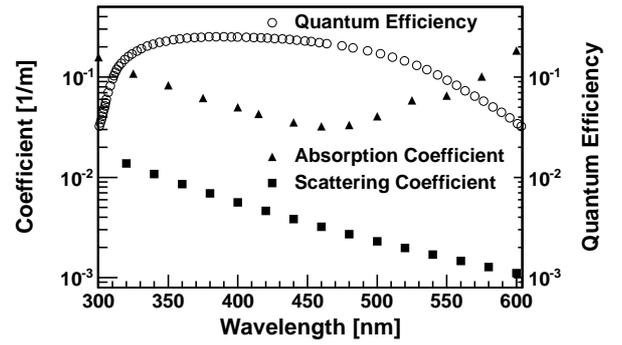}
  \caption{Absorption and scattering coefficient (reciprocal of
    the length) of the water transparency adopted in the simulation. The quantum
    efficiency of the PMT is also shown.}
  \label{fig:absmodel}
\end{figure}

The water absorption length of 405~nm is used to represent the water transparency. Values for other wavelengths are extrapolated from the curve for pure water
measured by Ref.~\cite{Querry1}. The coefficient 0.05 means that the attenuation length is 20 m at the specified wavelength 405~nm (Fig.~\ref{fig:absmodel}). The attenuation length mainly leads to absorption of the visible lights when the water transparency is not so good (e.g., attenuation length less than 40~m)~\cite{li2017}. This phenomenon indicates that the attenuation length is approximately equivalent to the absorption length in this study. For future experiment, 20~m attenuation length and effective water depth 4~m are reasonable running condition, so, those two parameters are adopted in the simulation. Fig.~\ref{fig:absmodel} is also shown that the typical PMT (R5912) photo-cathode quantum efficiency adopted in the simulation, according to the manufacturer's instruction manual~\cite{Hamamatsu}.

\section{Simulation results and experimental data}
The single-channel rate is related to primary cosmic rays. The flux of low energy cosmic rays and its secondary particles, especially the muon, n-p, and the EM components are very intense. These components produce Cherenkov light in the water, some of which may trigger the PMT.

\subsection{The single-channel rates of different components}\label{sect:MCrate}
\begin{figure}[htbp]
  \centering
  \includegraphics[width=0.9\linewidth]{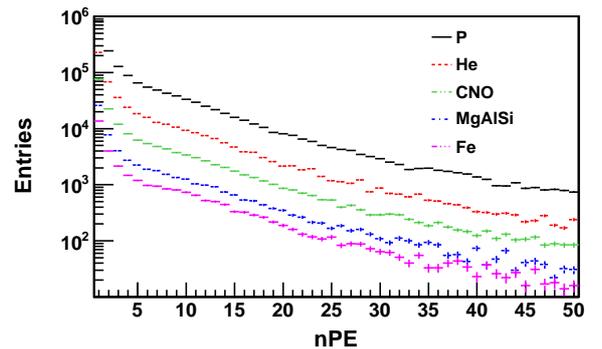}
  \caption{Trigger entries for the different primary CR components. In this simulation, the total time is 90~s. }
  \label{fig:znuclear}
\end{figure}

\begin{figure}[htbp]
  \centering
  \setlength{\abovecaptionskip}{0pt}
  \setlength{\belowcaptionskip}{0pt}
  \includegraphics[width=0.9\linewidth]{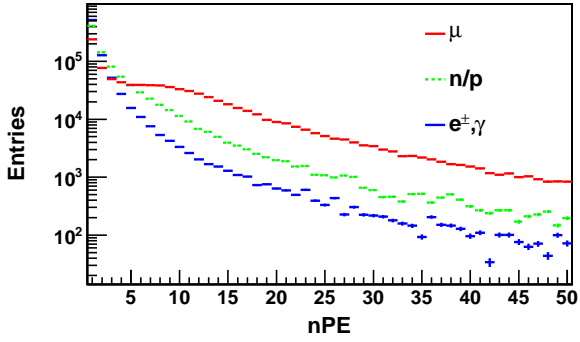}
  \caption{Charge distribution for different secondary particles of showers.}
  \label{fig:secparticle}
\end{figure}

Fig.~\ref{fig:znuclear} displays the trigger entries of different primary cosmic rays and time of simulation is 90 seconds. Single-channel rates are dominated by primary proton and helium, contributing to nearly 90\% of total single-channel rate. Other components totally share 10\%, as shown in Table~\ref{tab:rate}. Fig.~\ref{fig:secparticle} shows the charge distribution of secondary particles in the showers. In the same table, another set of data based on different secondary particles are also shown. This technique is used in experimental data analysis and correction, which will be described in Section~\ref{sect:ybjrate}. The contribution ratio of muons, n-p and EM to the single counting rate is roughly estimated as 10.3: 9.9: 8.8. The corrected MC total single-channel rate is approximately 31.4 kHz when the after-pulses rate (APR) is approximately 3.0\% and the PMT dark noise is approximately 1.5~kHz. The average rate of experimental data is within range 30.5 -- 32.2~kHz described in Section~\ref{sect:ybjrate}.

\begin{table}[htb]
  \centering
  \caption{The single-channel rate comparison.}
  \label{tab:rate}
  \small
  \begin{tabular}{lccccc}
    \hline
    primary CR & rate/kHz & &secondary particles & rate/kHz \\   
    \hline
    P       &  20.37  && $\mu$     & 10.27\\
    He      &  5.69   && n-p       & 9.91\\
    CNO     &  1.94   && $e^{\pm}$, $\gamma$ &  8.82 \\
    MgAlSi  &  0.68   &&           &      \\
    Fe      &  0.36   &&           &      \\
    total   & 29.04   && total     & 29.00\\
    \hline
    corrected& APR: 3.0\%  && dark noise: 1.5~kHz\\
    MC      & &&31.4~kHz \\
    \hline
    data    & &&30.5 -- 32.2~kHz\\
    \hline
  \end{tabular}
\end{table}

\subsection{Single-channel rate and water transparency}
\begin{figure}[htb]
  \setlength{\abovecaptionskip}{0pt}
  \setlength{\belowcaptionskip}{0pt}
  \centering
  \includegraphics[width=0.9\linewidth]{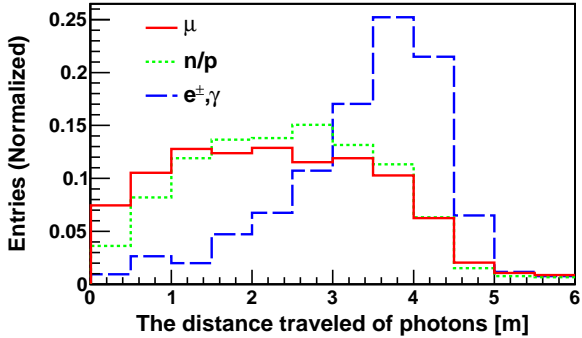}
  \caption{Distribution of the  distance traveled by the Cherenkov photons. }
  \label{fig:ddist}
\end{figure}

\begin{figure}[htb]
  \centering
  \includegraphics[width=0.9\linewidth]{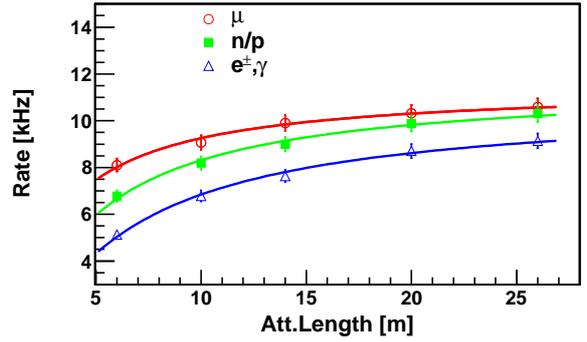}
  \caption{Single-channel rate for different secondary particles, from MC simulation, as a function of attenuation length. Different markers denote distinct secondary particles, as shown by the text in the plot. The curves are plotted using the simulation data points fitted by Equation ~(\ref{eq:rate3}).}
  \label{fig:zrate}
\end{figure}

Water transparency and single-channel rate are corrected to compare the simulation result and experiment data.
Fig.~\ref{fig:ddist} displays the distribution of distance traveled by Cherenkov photons from MC simulation. The photons caught by the PMT are generated at same distances, independent of the water transparency. Then, more photons are expected to reach the PMT when the water is more transparent, resulting in increased single-channel rate with the water attenuation length (Fig.~\ref{fig:zrate}).

Water transparency and single-channel rate obeys the exponential function, as follows:
\begin{eqnarray}
  \frac{R_{1}}{e^ {(-d/\lambda_1)}} =  \frac{R_{0}}{e^ {(-d/\lambda_0)}}.
  \label{eq:rate2}
\end{eqnarray}
$R_{\rm 1}$ is the single-channel rate when attenuation length of water is $\lambda_1$, $R_0$ is the single-channel rate when the water transparency is $\lambda_0$, and $d$ is defined as the distance  traveled by the Cherenkov photons. When the transparency $\lambda_0$ is infinitely well, Equation~(\ref{eq:rate2}) changes to Equation~(\ref{eq:rate3}), as follows:
\begin{eqnarray}
  R_{\rm 1} =R_{0} \cdot e^ {-d/\lambda_1}.
  \label{eq:rate3}
\end{eqnarray}

\subsection{Single-channel rate of prototype array}\label{sect:ybjrate}
In the prototype array, the single-channel rate in Section~\ref{sect:MCrate} can not be directly measured at a reasonable running condition (20~m attenuation length and effective water depth 4~m), because of imperfect sealing of pool and  the presence of strong radon radioactivity in the fresh water. The single-channel rate is approximately 45.5~kHz at attenuation length of 20 m, leads to a value of roughly 30.5~kHz after the radioactivity contribution is reduced to approximately 15.0~kHz.

The averaged single-channel rates for PMTs, at 34 m attenuation length in radon-contaminated water, are approximately 55.1~kHz, leading to a value of roughly 34.1~kHz after the radioactivity contribution is reduced (approximately 21.0~kHz). Thus, the single-channel rate at the reasonable running condition is estimated to be 32.2~kHz by the exponential law Equation~(\ref{eq:rate2}). At attenuation length of 7.5 m, the measured rate is 25.1~kHz in no-radon-contaminated water, which was then converted to 30.7~kHz at the reasonable running condition, approximately same as estimated above. Thus, the  estimated single-channel rates by these methods are in agreement with each other.

Therefore, at a reasonable running condition, we confirm the average rate value of experimental date within the range of 30.5 -- 32.2~kHz. The corrected MC total single-channel rate is approximately 31.4~kHz. Thus, at three different attenuation lengths, the MC single-channel rates are consistent with the real data in 3.5\% level (Table~\ref{tab:DATARATEbak}), which is well inside the real variation in time caused by the environmental changes.

\begin{table}[ht]
  \centering
  \caption{Comparison of the single-channel rates of experimental data with simulation results at different  attenuation lengths (Rate: kHz).}
  \label{tab:DATARATEbak}
  \small
  \begin{tabular}{lcccc}
    \hline
        & $\lambda$=20~m& $\lambda$=34~m& $\lambda$=7.5~m \\
    \hline
    data rate   & 45.5&55.1& 25.1  \\
    sinking&3.2\%/day& 4.5\%/day& -- \\
    radon radio  &15.0  &21.0& --\\
    \hline
    data correct   &30.5&34.1&25.1        \\
    MC correct     &31.4&35.3&24.4        \\
    difference       &2.9\%& 3.5\%&2.8\%    \\
    \hline
    data correct $\lambda$=20~m &30.5&32.2&30.7 \\
    \hline
  \end{tabular}
\end{table}

\subsection{Trigger rates at different PMT multiplicities}
\begin{figure}[htb]
  \centering
  \includegraphics[width=0.9\linewidth,clip]{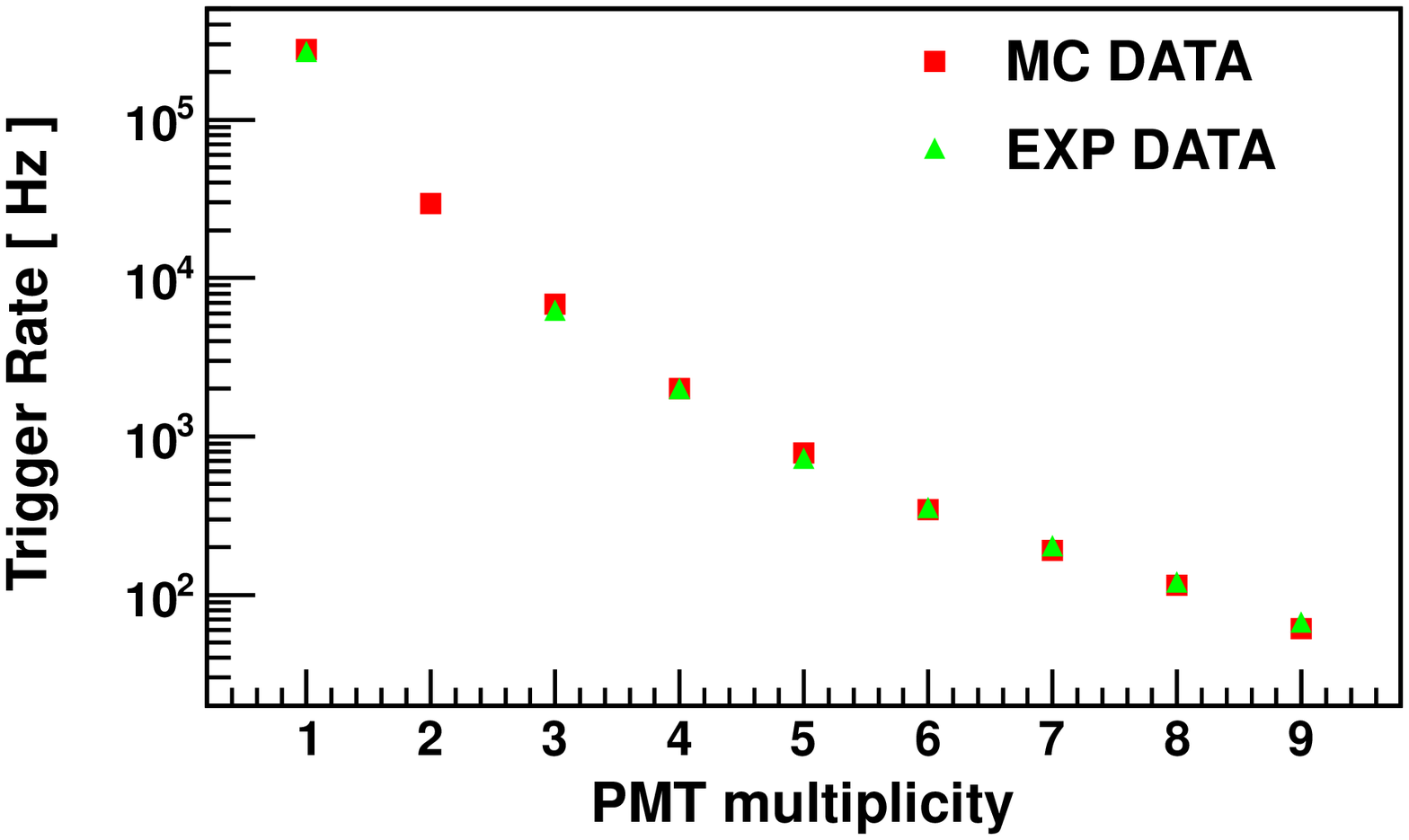}
  \includegraphics[width=0.9\linewidth,clip]{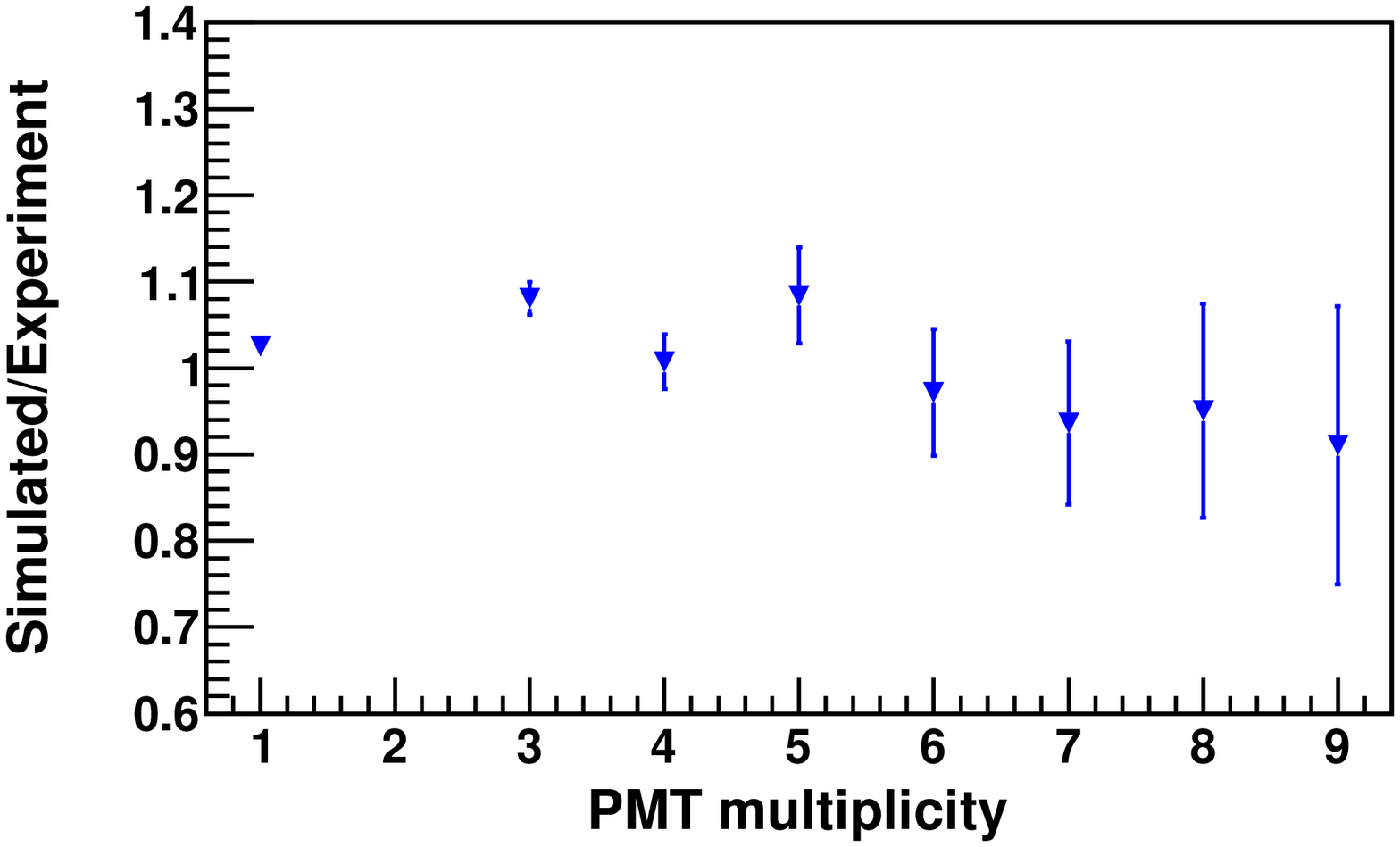}
  \caption{Trigger rate versus the PMT multiplicity for the simulation data and our experimental data (top) and the ratio of the trigger rate for the simulation data to the experimental data (bottom).}
  \label{fig:zztriger}
\end{figure}

The top of Fig.~\ref{fig:zztriger} shows the trigger rate  versus the PMT multiplicity for the simulation data and experimental data when the threshold is set to 1$/$3~PE for each PMT. The PMT multiplicity is the number of PMTs required to be in coincidence during a time window of 100~ns. The multiplicity requirement can be adjusted in a configuration file for the electronics system.
Furthermore, the ratio of the trigger rate for the simulation to the experimental at different PMT multiplicities is shown at the bottom of~\ref{fig:zztriger}. The trigger rate of the simulation data is consistency compared with the experimental data at 10\% level.

\subsection{Comparison on charge distribution}
Three peaks (at 1PE, 14PEs and 340PEs) are clearly observed in the analysis of the charge distribution of signals
originated from showers by MC simulation and experimental data.
Previous analyses~\cite{yao2011,an2011}
manifested that the first peak in the distribution comes from the
single photo-electron signals. The position of this peak can be used to monitor the gain changes of the PMT and the stability of the electronic system. The third peak originates from nearly vertical cosmic muons directly hitting the photo-cathode, which acts as a charge calibration parameter~\cite{gao2014}.
The second peak, which is formed by the pure geometrical effect of cosmic muons, is used to monitor and measure water transparency~\cite{li2017}. Analysis of the experimental data, that is, the single-channel rate of each PMT, at a threshold of roughly 8 PEs, shows the effective elimination of radon radio activity. This parameter can also be used to monitor and measure water transparency. More studies for this are in progress and will be discussed elsewhere.

\begin{figure}[htb]
  \centering \includegraphics[width=0.9\linewidth]{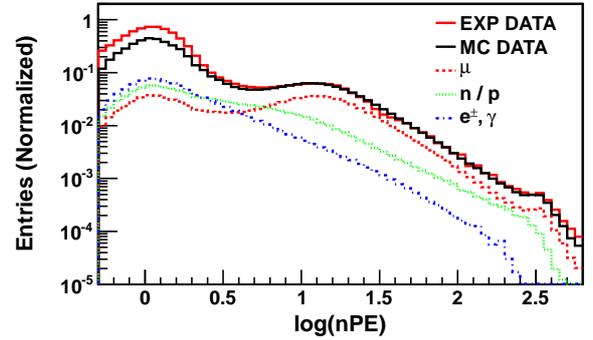}
  \caption{Charge distribution of experimental data (red solid line) and simulations (black solid line). Dotted line are the different components of showers from MC simulation: red, green, and blue lines represent the muon, n-p and EM, respectively, with scaling factors 1. The attenuation length of water is approximately 34~m.}
  \label{fig:chargedismc}
\end{figure}

The charge distribution of signals originated from MC simulation and experimental data are drawn in Fig.~\ref{fig:chargedismc}.
Three peaks in the charge distribution are obviously observed.
Simulation results are not in good agreement with experimental data under 3 PEs because of the effective radon radio activity. We don't know the charge distribution of radon radio, thus, the difference is not corrected. The charge distribution of the simulation data is little lower than the experimental data at high charge, most likely because the primary high-energy shower cannot be simulated.

\subsection{Discussions}
The injection area of grids is set $25~\rm m \times 25~\rm m$ in the simulation in section 3.1 because of the specific geometric structure of the WCDA prototype array. The effect of single-channel rate, contributed by muon particles out of this grid, is ignored because of the absorption of concrete walls and soil layer. However, for the LHAASO-WCDA, this effect, also known as the muon punch-through effect, should be considered. The other effect factor of single-channel rate is solar modulation. In the simulation, the fluxes of proton and helium are measured by AMS-02 between 2011 and 2013, corresponding with the functioning prototype array. In the future, the flux of low energy cosmic ray is highly different for longer periods. Thus, the solar modulation should be considered.

\subsubsection{Muon punch-through effect}
The cosmic muon performance is quite different from the EM components because the strong penetration by muons. Most muons can pass through more than one cell, yielding large signals, and simultaneously triggering many PMTs. Thus, the effective area of muon is much larger than one cell area 25~$\rm m^2$.

$A_{\rm eff}$ is the effective area of muon, as follows Equation~(\ref{eq:areaeff}):
\begin{eqnarray}
  A_{\rm eff} =  A_{\rm injection} \cdot \frac{N_{\rm trigger}}{N_{\rm injection}}
  \label{eq:areaeff}
\end{eqnarray}
where the $A_{\rm injection}$ is the area over which simulated events are thrown, $N_{\rm trigger}$ is the number of muons that result in at least the middle PMT being hit, and $N_{\rm injection}$ is the number of simulated muons.

\begin{figure}[htb]
  \centering
  \includegraphics[width=0.9\linewidth]{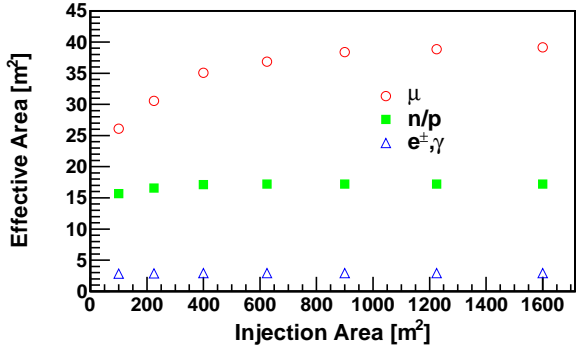}
  \caption{The effective area of the different secondary particles versus the injection area.}
  \label{fig:muoneffectivearea20}
\end{figure}

Fig.~\ref{fig:muoneffectivearea20} shows the effective area for muon as a function of the injection area. The effective area rapidly reaches the saturation value, the order of 36.9~$\rm m^2$  around 625~$\rm m^2$ injection area, and the rest remains nearly constant 39.1~$\rm m^2$ until the injection area reaches 1600~$\rm m^2$. Thus, the effective area of muon can be estimated to be 39~$\rm m^2$ and  an injection area overburden of $\approx$ 1225~$\rm m^2$ is enough for muon. Moreover, when the specific geometric structure of WCDA prototype array is considered, the injection area is set 625~$\rm m^2$  in above simulation, and the discordance is less than a level of 5.4\% for muon. Additionally, the same  analysis for other components of showers are observed. The effective area are constant 17.2~$\rm m^2$ (n-p) and 2.92~$\rm m^2$ (EM) when the injection area larger than 625~$\rm m^2$. Thus, the set of prototype array is enough to study the single-channel rate. The difference in the single-channel rates, because of the injection area from 625~$\rm m^2$ to 1225~$\rm m^2$, is approximately 1.8\% by considering the contribution ratio of the muon.

\subsubsection{Solar modulation effect}
Cosmic ray particles entering the heliosphere are scattered by irregularities in the heliospheric magnetic field. These particles experience convection and adiabatic deceleration in the expanding solar wind. The resulting modification of the cosmic ray energy spectra is known as {\sl solar modulation}.

\begin{figure}[htb]
  \centering
  \includegraphics[width=0.9\linewidth]{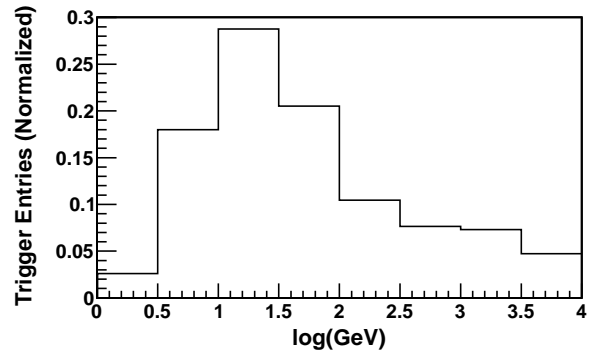}
  \caption{Energy distribution for triggered primary cosmic rays.}
  \label{fig:triggerEdis}
\end{figure}

The monthly average of proton fluxes, measured between 2011 and 2013 by AMS-02~\cite{Consolandi2015} in the first 30 months of operation, changed by the solar modulation 19\% (2.15 -- 2.40~GeV), 13\% (3.29 -- 3.64~GeV), 8\% (5.37 -- 5.90~GeV) and 3\% (10.10 -- 11.00~GeV). Considering the energy distribution for triggering primary cosmic rays shown in Fig.~\ref{fig:triggerEdis}, the uncertainty for the single-channel rate in solar modulation was finally found to be approximately 2\% below 10 GeV in this period.

The solar cycle includes both a sunspot activity period of approximately 11 years and approximately 22 year magnetic cycle with alternating positive and negative phases. Large differences at low energies have been observed in the proton and helium spectra measured by the BESS collaboration~\cite{Abe2016} and PAMELA~\cite{Adriani2013}. These differences reflect the variations in the solar activity and heliospheric magnetic field polarity. The annual average of primary proton fluxes can change by approximately 118.1\% at 2.51 GeV, 32.1\% at 5.62 GeV, and 8.1\% at 17.8~GeV, as shown in Table~\ref{tab:solarcycle}. The difference in the single-channel rates in solar modulation, in one solar cycle (sunspot activity period of 11 years), was found to be around 12\% by considering the energy distribution for primary cosmic rays.

\begin{table}[ht]
  \centering
  \caption{Difference of the annual average of proton fluxes between 1997 and 2009. And difference in the single-channel rate, considering the energy distribution for primary cosmic rays.}
  \label{tab:solarcycle}
  \small
  \begin{tabular}{lcccc}
    \hline
    energy range (GeV) & difference &  rate change\\
    \hline
    2.0  -- 3.16 (2.51)  & 118.1\% &3.01\% \\
    3.16 -- 10  (5.62)   & 32.1\%  &5.65\% \\
    10   -- 31.6 (17.8)  & 8.1\%   &2.28\% \\
    31.6 -- 100 (56.2)   & 2.3\%   &0.46\% \\
    \hline
    total                &    &$\approx$ 12\% \\
    \hline
  \end{tabular}
\end{table}

\section{Conclusion }

 In this study, the MC sample shows good consistency compared with the experimental data of the prototype array, which we used to observe the single-channel rate, PMTs hit multiplicities and the charge distribution. The average experimental of rate value is in the range of 30.5 -- 32.2 kHz after the radioactivity contribution is reduced. The corrected  MC total single-channel rate is approximately 31.4~kHz at 3\% level when the effect of after pulse, PMT dark noise and the water transparency are considered. The single-channel rate is dominated by proton and helium by contributing to nearly 90\%. For the secondary particles, the contribution ratio of hadrons to the single counting rate is approximately 34\%, and this parameter is an important component. Cosmic muons can pass through multiple cells, effective area (up to 36.9~$\rm m^2$) much larger than one cell area 25~$\rm m^2$, contributing to rate more than 35\%.

 Several realistic cases are considered. In the future, the single-channel rate will be increased to approximately 2\% by the muon punch-through effect because of the difference in the geometries between prototype array and future WCDA experimental pool. After a prolonged running period, the single-channel rate is deeply influenced by solar modulation, the difference in the single-channel rates in one solar cycle is approximately 6\%. These studies will be useful to fully comprehend future detectors.

\section*{Acknowledgments}
This work is partly supported by NSFC (No.11375224 and No.11675187); Joint Large-Scale Scientific Facility Funds of the NSFC and CAS under Contracts U1332201 and U1532258.

\end{document}